# Enhancing Grover's Search Algorithm: A Modified Approach to Increase the Probability of Good States

Ismael Abdulrahman, ismael.abdulrahman@epu.edu.iq
Department of Technical Information Systems Engineering, Erbil Technical Engineering College, Erbil Polytechnic University, Erbil 44001, Kurdistan region–Iraq.

*Abstract*—This article introduces an enhancement to the Grover search algorithm to speed up computing the probability of finding good states. It suggests incorporating a rotation phase angle determined mathematically from the derivative of the model during the initial iteration. At each iteration, a new phase angle is computed and used in a rotation gate around $y + z$ axis in the diffusion operator. The computed phase angles are optimized through an adaptive adjustment based on the estimated increasing ratio of the consecutive amplitudes. The findings indicate an average decrease of 28% in the required number of iterations resulting in a faster overall process and fewer number of quantum gates. For large search space, this improvement rises to 29.58%. Given the computational capabilities of the computer utilized for the simulation, the approach is applied to instances with up to 12 qubits or 4096 possible combination of search entries.

*Index Terms*— quantum computing, Grover search algorithm, adaptive search, optimization.

## 1.1 Introduction

Classical search algorithms, such as the brute-force approach, exhibit a time complexity proportional to the size of the search space—$O(N)$, where $N$ represents the number of possible solutions. Grover's algorithm, on the other hand, achieves a quadratic speedup, reducing the time complexity to $O(\sqrt{N})$ [1–3].

Grover's algorithm revolves around two key quantum computing principles: superposition and interference. It exploits the quantum parallelism inherent in superposition to evaluate multiple possibilities simultaneously. Additionally, interference is strategically harnessed to amplify the probability amplitude of the correct solution, while diminishing the amplitudes of incorrect ones.

The algorithm begins with the creation of a superposition of all possible states representing the search space. Subsequently, a quantum oracle is employed to mark the target state, effectively inverting its amplitude. The algorithm then utilizes a series of quantum operations, including amplitude amplification, to boost the probability of measuring the correct solution. This process is iteratively applied, leading to a quadratic speedup in finding the desired result compared to classical algorithms.

The Grover search algorithm has been thoroughly described in various references and sources, including those mentioned in [1–10]. The description presented in [7] is adopted in this paper. For a more profound understanding and additional insights into the algorithm, one can refer to the aforementioned references.

Several modifications are proposed in the literature, including circuit decomposition using unitary matrices [11], elimination of the diffuser's gate by replacing the Hadamard gate with $R^X_{\pi/2}$ [12], the binomial version of the Grover algorithm which reduces the number of iterations but increases the number of gates required for the search [13], and the use of Clifford's geometric algebra to visualize the search process as a spin-1/2 particle [14]. In [15], a method was introduced requiring only O(ln(N)) iterations in certain cases to locate the target state. However, achieving consistent improvement in discovering the marked state proved challenging, which limits the practical utility of the proposed method. Reference [16] demonstrated the possibility of accelerating the search by dividing the register, albeit at the expense of increasing the algorithm's complexity. In [17], it was shown that improving the algorithm is feasible when there's an imbalance in the counts of 0s and 1s. However, this improvement doesn't necessarily boost computational efficiency; rather, it simplifies implementation. In [18], authors proposed a modified version requiring fewer gates, resulting in a 12% improvement in accuracy and a 21% decrease in execution time compared to the original algorithm. Lastly, in [19], authors suggested a variational approach resulting in slight performance enhancements for different qubit configurations, with improvements of 5.77% and 3.95% for three and four qubits, respectively. In section 1.3, an adaptive adjustment is proposed to increase the chance of finding the good state by reducing the required number of iterations by 28%. For large-scale search space, this improvement rises to 29.5%. The process can be implemented with fewer number of quantum gates. The proposed approach is applied to models with up to 12 qubits or equivalent 4096 possible combination of search entries. In addition, another similar approach is presented that include different formulas for computing the phase angles.

## 1.2 The Algorithm

Grover's algorithm was first introduced by Grover in 1996 [20]. It addresses the challenge of finding a solution $x_0$ such that $f(x_0) = 1$ mapping the function $f(x): \{0,1\}^n \to \{0,1\}$, where $n$ denotes the bit-length of the search space. The algorithm's complexity is determined by how often the function $f(x)$ is called. In the worst-case scenarios of classical methods such as the brute-force algorithms, the function needs to be called $N - 1$ times where $N = 2^n$, covering all potential options in the search space. Grover's quantum algorithm notably speeds up this procedure, achieving a quadratic acceleration. Here, "quadratic" indicates that only around $\sqrt{N}$ evaluations are needed, in contrast to the classical requirement of $N = 2^n$.

Consider $N = 2^n$ eligible items for a search task indexed with integers from 1 to $N − 1$, and $M$ distinct inputs for which $f(x) = 1$. The algorithm follows these steps [7]:
1. Create a register of $n$ qubits set initially to the state $|0\rangle$.
2. Apply H gate to each qubit of the register to prepare them for balanced superposition using the formula $\frac{1}{\sqrt{N}}\sum_{x=0}^{N-1}|x\rangle$, where $\frac{1}{\sqrt{N}}$ refers to the uniform amplitude for each state $|x\rangle$.
3. Apply the following 4 steps repeated $N_{optimal}$ times where $N_{optimal} = \frac{\pi}{4}\sqrt{\frac{N}{M}} - \frac{1}{2}$:
   a. Mark the good state or solution using the phase oracle $O_f$ that applies a negative sign to that target state.
   b. Apply H gate to each qubit in the register.
   c. Change the sign of every computational basis state except $|0\rangle$.
   d. Apply H gate again as in 3b. We call the above 3b, c, and d steps as Grover discussion operator.
4. Apply measurement to the register to highlight the state index with high probability.
5. Return to step 3 if the condition is not met

The complete unitary operation applied to the register can be expressed concisely in a single equation:

$$\left(\boxed{-H^{\otimes n}O_0H^{\otimes n}}\, O_f\right)^{N_{optimal}} H^{\otimes n}|0\rangle \quad (1)$$

The green-highlighted text is enclosed within a frame to emphasize that this section constitutes the paper's contribution to the algorithm which will be described in the following sections. In (1), the blue shading represents the preparation stage (step 2), the red shading corresponds to the oracle step (step 3a), and the framed green shading denotes the Grover diffusion operator (step 3b–d) which is along with the oracle, are iterated according to the power specified in the equation ($N_{optimal}$ times).

Note that (1) can be rewritten after substituting $O_0 = X^{\otimes n}(c^{n-1}Z)X^{\otimes n}$ as follows:

$$\left(\boxed{-H^{\otimes n}X^{\otimes n}(c^{n-1}Z)X^{\otimes n}H^{\otimes n}}\, O_f\right)^{N_{optimal}} H^{\otimes n}|0\rangle \quad (2)$$

In (2) we have $(n-1)$-fold-controlled-Z sandwiched between $X^{\otimes n}H^{\otimes n}$. The process occurs in a right-to-left manner, involving preparation first, followed by phase flip, and concluding with amplitude amplification (blue, red, then green).

### 1.3 The proposed approach

The Z gate mentioned in (2) functions as a phase-flipping operator that causes a half-cycle phase change in the qubit state, leveraging the relationship $e^{\pi i} = -1$. As a result, the inversion-about-the-mean operation can be interpreted in two ways: either by inverting the amplitude of the specific state intended for amplification as in the standard algorithm, or by considering it as a desired phase rotation that doesn't have to be $\pi$, and our goal is to augment it.

This prompts the following question:
*Is it possible to increase the probability of success by incorporating the increasing ratio of amplitudes in each iteration within the diffusion operator using a phase-angle rotation gate?*

We need to estimate this ratio and hence determine the appropriate phase angle to employ in the operator. By examining the Z gate closely, we observe that it does not change the amplitude of the state from a real to a complex value or vice versa (rotation by $\pi$). Similarly, we understand that the rotation around y-axis by any value has the same effect; in other words, if the amplitudes have real values, rotating them around the y-axis preserves these values as real. Such preservation of real values cannot be accomplished through rotations around the x- or z-axis except for the cases where the rotation angle is $\pi$.
Hence, our interest lies in investigating whether the rotation by $\pi$ applied by the Z gate in the diffusion operator (2) results in the maximum amplitude following the reflection around the mean.
The rotation around the y-axis is described by the following equation:

$$R_\theta^Y = \begin{bmatrix} \cos\frac{\theta}{2} & -\sin\frac{\theta}{2} \\ \sin\frac{\theta}{2} & \cos\frac{\theta}{2} \end{bmatrix} \quad (3)$$

Multiplying (3) by the matrix of Z-gate we get $R_{\theta\pi}^{YZ}$ which is a rotation around z-axis by $\pi$ and y-axis by the angle $\theta$:

$$R_{\theta\pi}^{YZ} = R_\theta^Y R_\pi^Z = \begin{bmatrix} \cos\frac{\theta}{2} & -\sin\frac{\theta}{2} \\ \sin\frac{\theta}{2} & \cos\frac{\theta}{2} \end{bmatrix}\begin{bmatrix} 1 & 0 \\ 0 & -1 \end{bmatrix} = \begin{bmatrix} \cos\frac{\theta}{2} & \sin\frac{\theta}{2} \\ \sin\frac{\theta}{2} & -\cos\frac{\theta}{2} \end{bmatrix} \quad (4)$$

Now, our objective is to determine the phase angle $\theta$ that results in a maximum amplitude in the diffusion operator stage. This can be achieved by taking the derivative of (2) with respect to the angle $\theta$ and setting it equal to zero:

$$\frac{d}{d\theta}\left(\boxed{-H^{\otimes n}X^{\otimes n}\left(c^{n-1}(ZR_\theta^Y)\right)X^{\otimes n}H^{\otimes n}}\, O_f H^{\otimes n}|0\rangle\right) = 0 \quad (5)$$

$$\frac{d}{d\theta}\left(\boxed{-H^{\otimes n}X^{\otimes n}\left(c^{n-1}\begin{bmatrix}\cos\frac{\theta}{2} & \sin\frac{\theta}{2}\\ \sin\frac{\theta}{2} & -\cos\frac{\theta}{2}\end{bmatrix}\right)X^{\otimes n}H^{\otimes n}}\, O_f H^{\otimes n}|0\rangle\right) = 0 \quad (6)$$

Notice how (6) can be restated by integrating the operations of the Hadamard and Pauli-X gates as follows:

$$\frac{d}{d\theta}\left(\boxed{-R^{\otimes n}\left(c^{n-1}\begin{bmatrix}\cos\frac{\theta}{2} & \sin\frac{\theta}{2}\\ \sin\frac{\theta}{2} & -\cos\frac{\theta}{2}\end{bmatrix}\right)R^{\dagger\otimes n}}\, O_f H^{\otimes n}|0\rangle\right) = 0 \quad (7)$$

where

$$R = HX = \frac{1}{\sqrt{2}}\begin{bmatrix} 1 & 1 \\ 1 & -1 \end{bmatrix}\begin{bmatrix} 0 & 1 \\ 1 & 0 \end{bmatrix} = \begin{bmatrix} \frac{1}{\sqrt{2}} & \frac{1}{\sqrt{2}} \\ -\frac{1}{\sqrt{2}} & \frac{1}{\sqrt{2}} \end{bmatrix}$$

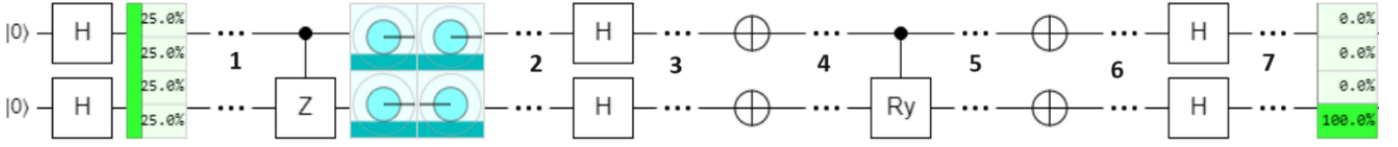

Figure 1 Modified Grover search algorithm for $n = 2$ qubits, showing the substitution of the Z gate with the gate $R_{\theta\pi}^{YZ}$. The subsequent analysis breaks down each highlighted step above.

$$R = \begin{bmatrix} \cos\frac{-\pi}{4} & -\sin\frac{-\pi}{4} \\ \sin\frac{-\pi}{4} & \cos\frac{-\pi}{4} \end{bmatrix} = R_{-\pi/2}^Y$$

Similarly,

$$R^\dagger = (HX)^\dagger = \begin{bmatrix} \cos\frac{-\pi}{4} & -\sin\frac{-\pi}{4} \\ \sin\frac{-\pi}{4} & \cos\frac{-\pi}{4} \end{bmatrix}^\dagger = \begin{bmatrix} \cos\frac{\pi}{4} & -\sin\frac{\pi}{4} \\ \sin\frac{\pi}{4} & \cos\frac{\pi}{4} \end{bmatrix} = R_{\pi/2}^Y$$

Therefore, (6) can be rewritten as follows:

$$\frac{d}{d\theta}\left( -R_{-\pi/2}^{Y^{\otimes n}} \left( c^{n-1} \begin{bmatrix} \cos\frac{\theta}{2} & \sin\frac{\theta}{2} \\ \sin\frac{\theta}{2} & -\cos\frac{\theta}{2} \end{bmatrix} \right) R_{\pi/2}^{Y^{\otimes n}} O_f H^{\otimes n}|0\rangle \right) = 0 \quad (8)$$

Equation (6) can be solved for different values of $n$. A program in MATLAB is developed to give us the results of the angle $\theta$ so that we get maximum amplitudes after reflection about the mean. Let's Initially begin by considering the case where $n = 2$. Quirk simulator from IBM is used for simulating the results using a gate-block diagram as in Fig. 1.

The output of each step is as follows:
Step 1 (initialization to create balanced superposition):
$\frac{1}{2}(|00\rangle + |01\rangle + |10\rangle + |11\rangle)$
Step 2 (marking the target using the oracle's phase flip):
$\frac{1}{2}(|00\rangle + |01\rangle + |10\rangle - |11\rangle)$
Step 3 (remains unchanged as before):
$\frac{1}{2}(|00\rangle + |01\rangle + |10\rangle - |11\rangle)$
Step 4:
$\frac{1}{2}(-|00\rangle + |01\rangle + |10\rangle + |11\rangle)$
Step 5:
$\frac{1}{2}\left(-|00\rangle + |01\rangle + \left(\cos\frac{\theta}{2} + \sin\frac{\theta}{2}\right)|10\rangle + \left(\sin\frac{\theta}{2} - \cos\frac{\theta}{2}\right)|11\rangle\right)$
The optimal angle is found from (6) to be zero for this 2-qubit example. Therefore, the output of step 5 becomes:
$\frac{1}{2}(-|00\rangle + |01\rangle + |10\rangle - |11\rangle)$

Step 6:
$\frac{1}{2}\left(\left(\sin\frac{\theta}{2} - \cos\frac{\theta}{2}\right)|00\rangle + \left(\cos\frac{\theta}{2} + \sin\frac{\theta}{2}\right)|01\rangle + |10\rangle - |11\rangle\right)$
Substituting $\theta = 0$, the output becomes:
$\frac{1}{2}(-|00\rangle + |01\rangle + |10\rangle - |11\rangle)$
Which is the same result from step 5.

Step 7:
$\frac{1}{2}\left(\sin\frac{\theta}{2}|00\rangle + \left(1 - \cos\frac{\theta}{2}\right)|01\rangle + \sin\frac{\theta}{2}|10\rangle + \left(-\cos\frac{\theta}{2} - 1\right)|11\rangle\right)$
Which results to $-|11\rangle$ with a probability 100% and a single iteration ($N_{optimal} = 1$).

Since the optimal angle for this example was found to be zero and the number of iterations for the probability of success is only one, we noticed no impact of this rotation on speeding up the process. Let's move to scenarios involving higher number of qubits requiring additional iterations, employing the proposed approach.

Solving (6) for $n = 2, 3, 4, 5, 6, 7$, we obtain the following corresponding phases shown in Table I.

We can readily derive a general formula for the phase angle by examining Table I, yielding:

$$\theta = 2\tan^{-1}\left(\frac{2^{n-2}-1}{2^{n-2}}\right) = 2\tan^{-1}\left(\frac{N-4}{N}\right) = 2\tan^{-1}\left(1 - \frac{4}{N}\right) \quad (9)$$

For the case where $n = 2$ in (9), the result is $\theta = 0$, explaining the attainment of a 100% probability of locating the target state in a single iteration (Classically, we need four iterations for this problem in the worst-case scenario). However, the standard Grover diffusion operator assumes $\theta = 0$ for all iterations and for any number of qubits, which is found not to be optimal. It should be noted that solving (6) for $n$ greater than 12 qubits demands significant memory resources and entails a slow processing speed. Another insight observed from (9) is that, as $N$ increases, the phase angle $\theta$ iteratively approaches a phase angle of $\frac{\pi}{2}$.

Table I Phase angles corresponding to maximum probability

| Number of qubits ($n$) | 2 | 3 | 4 | 5 | 6 | 7 |
|---|---|---|---|---|---|---|
| Phase angle ($\theta$) | 0 | $2\tan^{-1}\frac{1}{2}$ | $2\tan^{-1}\frac{3}{4}$ | $2\tan^{-1}\frac{7}{8}$ | $2\tan^{-1}\frac{15}{16}$ | $2\tan^{-1}\frac{31}{32}$ |

Nevertheless, equation (9) does not provide information regarding the change in phase with each iteration. We want to examine how the inclusion of amplitude amplification affects this equation. To accomplish this, the ratio between two amplitudes in every two successive iterations is computed. In the initial iteration, the amplitudes of all states, including the target state $a_1$, are identical, given that they are balanced uniform states. Consequently, the mean $m_1$ equals the amplitude of the target state, i.e., $m_1 = a_1 = \frac{1}{\sqrt{N}}$.

In the second iteration, we initially compute the newly amplified amplitude resulting from the reflection about the mean $m_1$. This gives us the amplitude $m_2$ and $a_2$:
$m_2 = \frac{N-2}{N\sqrt{N}} = \frac{1}{\sqrt{N}} - \frac{2}{N\sqrt{N}}$ and $a_2 = \frac{3N-4}{N(\sqrt{N})} = \frac{3}{\sqrt{N}} - \frac{4}{N\sqrt{N}}$

Proceeding in calculating the updated mean following the phase reversal and the new amplitude after reflecting about the mean, the following are formulas of amplitudes for iterations to 1–7:

Table II Result of amplitude amplification at each iteration and the ratio between two successive amplitudes

| Iteration ($i$) | Amplitude ($a_i$) | $\sim \left(\frac{a_{i+1}}{a_i}\right)$ |
|---|---|---|
| 1 | $\frac{1}{\sqrt{N}}$ | - |
| 2 | $\frac{3}{\sqrt{N}} - \frac{4}{N\sqrt{N}}$ | $3$ |
| 3 | $\frac{5}{\sqrt{N}} - \frac{20}{N\sqrt{N}} + \frac{16}{N^2\sqrt{N}}$ | $\frac{5}{3}$ |
| 4 | $\frac{7}{\sqrt{N}} - \frac{56}{N\sqrt{N}} + \frac{112}{N^2\sqrt{N}} - \frac{64}{N^3\sqrt{N}}$ | $\frac{7}{5}$ |
| 5 | $\frac{9}{\sqrt{N}} - \frac{120}{N\sqrt{N}} + \frac{432}{N^2\sqrt{N}} - \frac{576}{N^3\sqrt{N}} + \frac{256}{N^4\sqrt{N}}$ | $\frac{9}{7}$ |
| 6 | $\frac{11}{\sqrt{N}} - \frac{220}{N\sqrt{N}} + \frac{1232}{N^2\sqrt{N}} - \frac{2816}{N^3\sqrt{N}} + \frac{2816}{N^4\sqrt{N}} - \frac{1024}{N^5\sqrt{N}}$ | $\frac{11}{9}$ |
| 7 | $\frac{13}{\sqrt{N}} - \frac{364}{N\sqrt{N}} + \frac{2912}{N^2\sqrt{N}} - \frac{9984}{N^3\sqrt{N}} + \frac{16640}{N^4\sqrt{N}} - \frac{13312}{N^5\sqrt{N}} + \frac{4096}{N^6\sqrt{N}}$ | $\frac{13}{11}$ |

We can observe an elegant pattern linking each amplitude to the preceding and following amplitudes. The first term of all amplitudes increases by the amount $(2i + 1)$ where $i$ refers to the iteration index. The terms exhibit alternate changes in signs, similar to the sine and cosine expansions while the denominators show a rapid exponential decrease. The impact of the other terms is negligible and, therefore, can be disregarded, leaving only the first part participating into the amplitude. Observe also the proportion between every successive pair of amplitudes, expressed as $\left(\frac{2i+1}{2i-1}\right)$ which is proportional to the ratio $\left(\frac{\sin\theta_{i+1}}{\sin\theta_i}\right)$. This augment can further be improved by reversing the indices added to unity, that is $\left(1 + \frac{2i-1}{2i+1}\right)$. The ratio provides an approximate relationship between the increment in phase angles during each two consecutive iterations. Observe as well that this rate is not uniform but declines as iterations progress (see the last column in Table II). This highlights the significance of the initial iterations, as they bear more importance compared to the subsequent ones, which remain relatively constant. As a result, this suggests to integrate this value into the ultimate rotation gate ($R^{YZ}_{\theta\pi}$–gate matrix). Therefore, (9) becomes:

$$\theta_i = \left(2\tan^{-1}\left(\frac{2^{n-2}-1}{2^{n-2}}\right)\right) + 1 + \frac{2i-1}{2i+1} \quad (10)$$

Alternatively, we can revise equation (10) so that it remains independent of the iteration number. This means that the phase angle remains constant, determined solely by the number of qubits $(n)$ and is not influenced by the iteration index $(i)$ as denoted by (9).

For the first iteration of the algorithm, we employ equation (11), while for the remaining iterations, we use equation (12):

$$-H^{\otimes n}X^{\otimes n}\left(c^{n-1}\left(ZR^Y_\theta\right)\right)X^{\otimes n}H^{\otimes n}O_f H^{\otimes n}|0\rangle \quad (11)$$

$$-H^{\otimes n}X^{\otimes n}\left(c^{n-1}\left(HR^Y_\theta\right)\right)X^{\otimes n}H^{\otimes n}O_f \quad (12)$$

In the context of equations (11) and (12), the phase angle denoted by $\theta$, calculated from equation (9), is smaller than the $\theta$ value in equation (10). This variation is adjusted by the additional rotation to the state through applying a Hadamard gate in (12) which represents a rotation around the $x + z$ axis. Specifically, for the first equation, we incorporate the $ZR^Y_\theta$ gates into the operator, whereas for the second equation, we employ the gates $HR^Y_\theta$. The proposed approach is used to simulate the GSA problem and presented in the next section. All simulation programs are coded in MATLAB and are provided with this paper.

### 1.4 Simulation Results:

In the Quirk toolbox by IBM, the Grover algorithm example employs a 5-qubit, 32-state configuration to illustrate how the standard algorithm functions in the search process. For comparison with the proposed method, the same example is used. While the standard algorithm requires 4 iterations, the proposed method achieves an equivalent result with only 3 iterations, representing a 25% reduction compared to the standard approach. The simulation is presented in Fig. 2, while the example from Quirk can be accessed on the provided Quirk's website. Observe the distinction between the proposed method and the standard algorithm, as the proposed approach introduces a mean distributed unevenly among the non-solution states for all iterations (see the green narrow column in Fig. 2). For this example, the probability of finding the good state using the proposed approach (here the state $|11111\rangle$) is 99.7461% versus 89.6936% for the standard algorithm.

Figure 3 presents comparison results for additional scenarios, considering number of qubits up to 13. The optimal number of iterations for the modified version denoted by $N_{optimal}$ is highlighted in red. The proposed method leads to an average decrease in the required number of iterations around 28% whereas for high search space, this improvement rises to 29.58% (Table III). It should be noted that in scenarios involving a significant number of qubits ($n$ greater than 12), advanced CPU and GPU resources are required to simulate the problem.

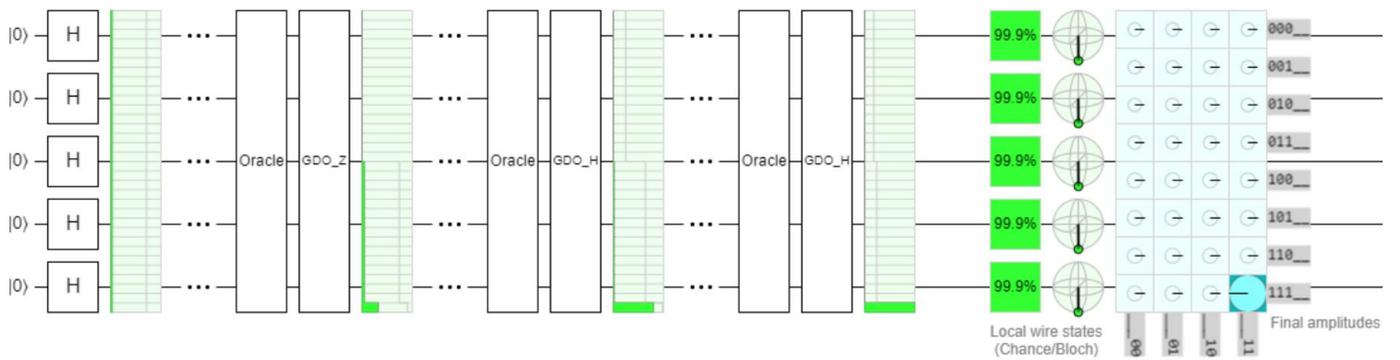

Figure 2 Simulation of the modified Grover search algorithm for $n = 5$ qubits employing equations (11)–(12)

Table III Comparison between the proposed and standard GSA

| Number of qubits (n) | 2 | 3 | 4 | 5 | 6 | 7 | 8 | 9 | 10 | 11 | 12 | 13 |
|---|---|---|---|---|---|---|---|---|---|---|---|---|
| Standard GSA | 1 | 2 | 3 | 4 | 6 | 8 | 12 | 17 | 25 | 35 | 50 | 71 |
| Modified GSA | 1 | 1 | 2 | 3 | 4 | 6 | 9 | 12 | 18 | 25 | 35 | 50 |
| Difference | 0 | 1 | 1 | 1 | 2 | 2 | 3 | 5 | 7 | 10 | 15 | 21 |
| Ratio | 1 | 0.5 | 0.6667 | 0.75 | 0.6667 | 0.75 | 0.75 | 0.7059 | 0.72 | 0.7143 | 0.70 | 0.7042 |
| Improvement | 0% | 50% | 33.33% | 25% | 33.33% | 25% | 25% | 29.41% | 28% | 28.57% | 30% | 29.58% |

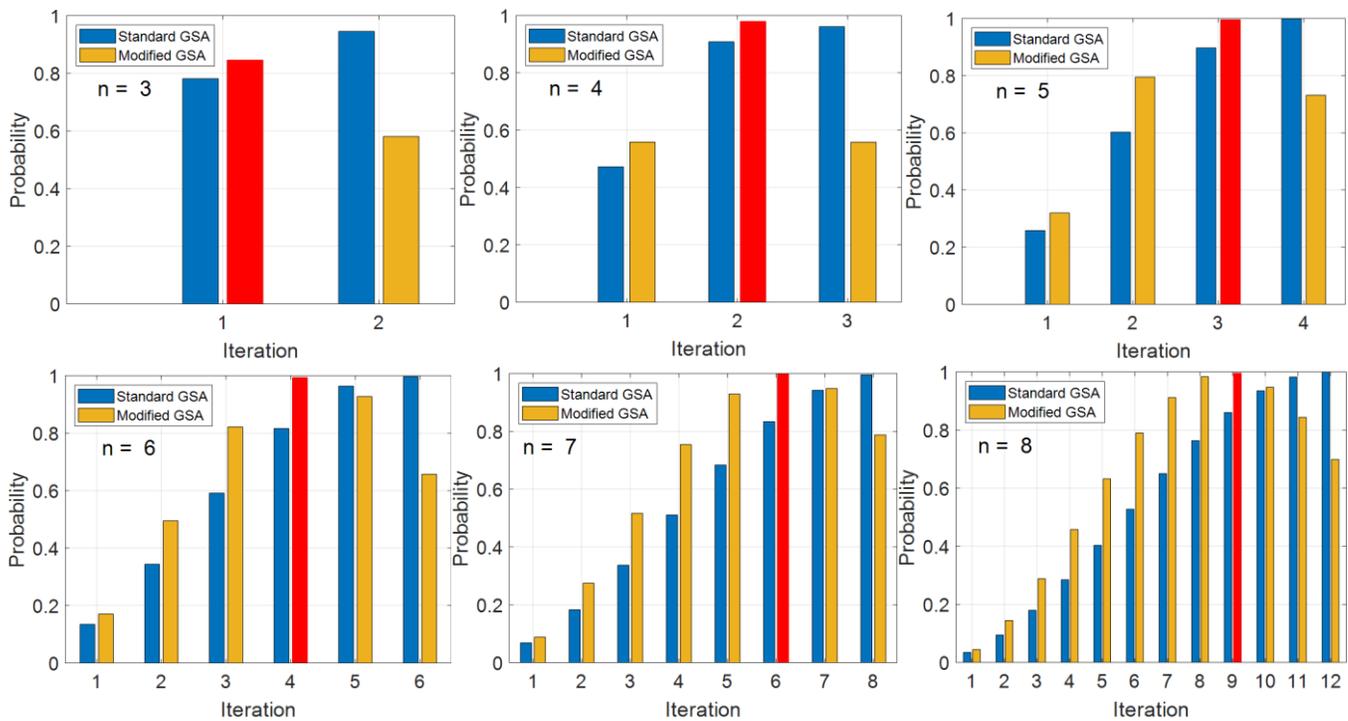

Figure 3 Comparison between the standard and modified Grover search algorithm for $n = 3, 4, 5, 6, 7, 8$. Red color is used to highlight the iteration that yields the maximum amplitude.

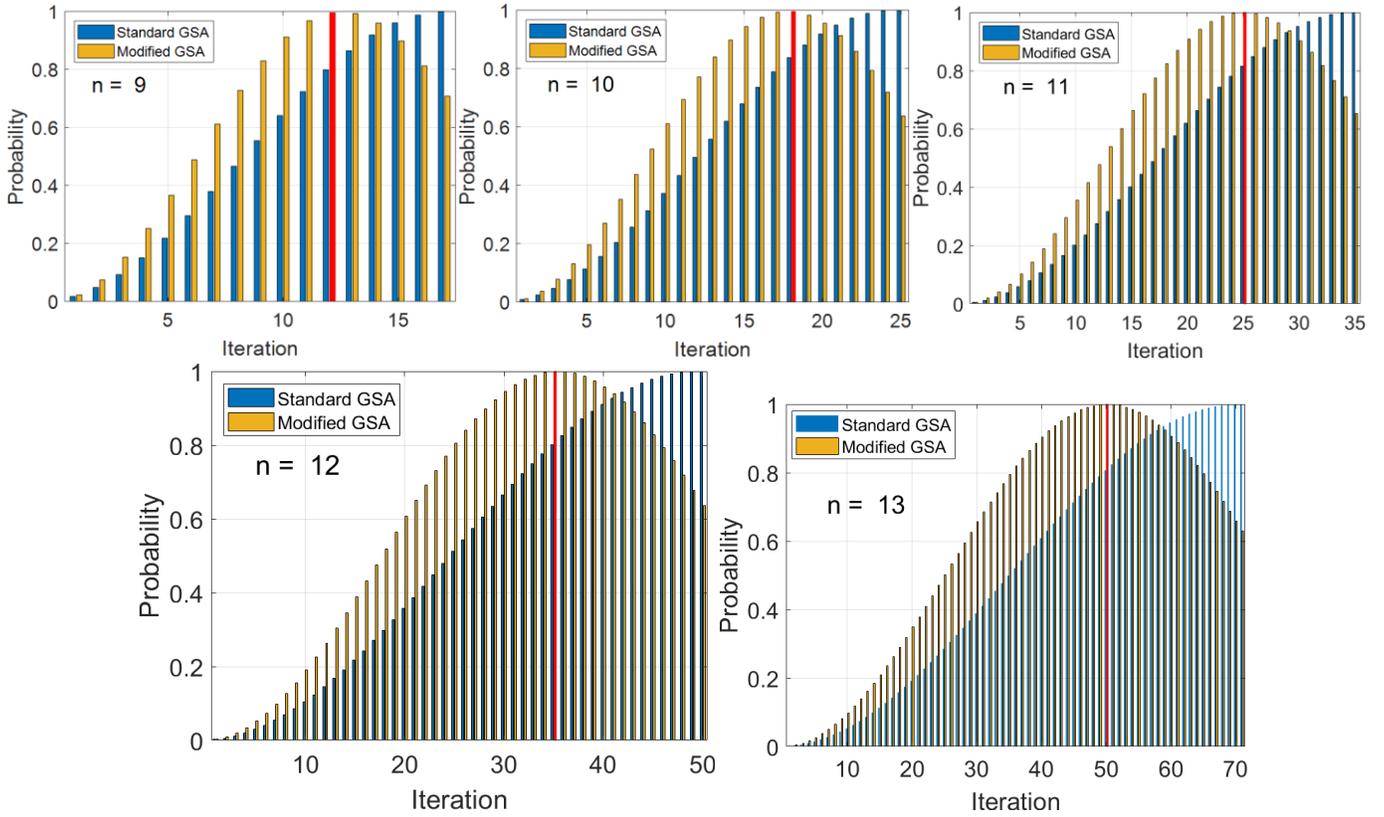

Figure 3 Comparison between the standard and modified Grover search algorithm for $n = 9, 10, 11, 12, 13$. Red color is used to highlight the iteration that yields the maximum amplitude.

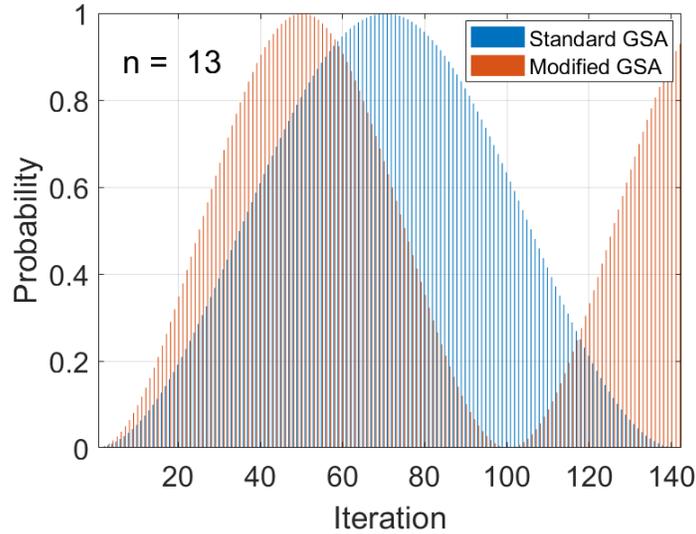

Figure 4 Comparison between the standard and modified Grover search algorithm for $n = 13$ lasting for 140 iterations

Precisely, the probability of success in the modified GSA's equation exhibits a 28.1017% average reduction of number of iterations (corresponding to a 1.42 ratio in the phase angle) reaching optimality in probability of success with only $0.72\, N_{optimal}$ iterations. Excluding the basic two-qubit scenario from this calculation, which is a special case involving only one iteration, results in an average enhancement by 30.65% in the number of iterations.

The original probability-of-success equation for the standard GSA is [7]:

$$P(success) = sin^2((2i+1)\theta) \qquad (7)$$

where $i$ refers to iteration index. The modified approach adjusts the above equation to the following:

$$P(success) = sin^2\big((2i+1)((1+\Delta\theta)\theta)\big) \qquad (8)$$

Here, $(1+\Delta\theta)$ represents the compression ratio of the probability function (the squared sine) or, in other words, the increasing ratio of the number of iterations, which is approximately $\sqrt{2}$. This observation is understandable as the additional phase angle contributes to a decrease in the number of iterations required for this problem to attain its maximum probability evidenced in Fig. 4. The probability of success represented by the $y$–axis increases as number of iterations increases (represented by the $x$–axis) until it reaches its optimal value $N_{optimal}$.

Concerning the number of gates needed by the proposed method, there is no extra gate required compared to the standard approach. In fact, implementing equation (8) instead of equation (6) results in a lower number of gates needed which is another improvement provided by this approach.

It should be noted that despite this improvement in the speed of success offered by the proposed approach, it does not result in a significant reduction in the algorithm complexity determined by the number of times the oracle is called. In the standard algorithm, its complexity is represented by $\left(\frac{\pi}{4}\sqrt{\frac{N}{M}} - \frac{1}{2}\right)$ compared to $\frac{1}{\sqrt{2}}\left(\frac{\pi}{4}\sqrt{\frac{N}{M}} - \frac{1}{2}\right)$ for the modified algorithm which is still considered a quadratic speedup as in the standard algorithm, denoted by $O(\sqrt{N})$.

## 1.5 Conclusion

This study introduced an improved version of the Grover search algorithm that aimed at maximizing the amplitudes, and consequently the probability, of desired states known as good, target or solution states. The proposed modification includes integrating both the amplitude-increasing ratio of consecutive iterations and the derivative of the Grover diffusion operator. This combination is employed to determine the optimal phase angle used for amplitude amplification by utilizing a rotation around the $y + z$–axis. The findings illustrate an average decrease in the required number of iterations to achieve a probability of success around 28% less compared to the standard Grover search algorithm. For large-scale search space, this improvement rises to 29.58%. An additional alternative, which is similar in nature, is also illustrated by equations (11) and (12), utilizing a fixed phase angle within the rotation gate around the y-axis. The proposed approach can be implemented with fewer gates if we choose to apply equation (8) instead of equation (6) which is used for the standard diffusion. The research includes various case studies, including scenarios with up to 12 qubits resulting in 4096 combinations of search entries. The simulation was conducted using MATLAB and IBM's Quirk programs.

## Funding
No financial support is provided for this work.

## Conflict of Interest
The authors declare that there is no conflict of interest associated with this work.

## Code and Data Availability
The author would like to share the MATLAB program developed exclusively for this paper with the readers for public access. No specific data is used for this research.
https://www.mathworks.com/matlabcentral/fileexchange/158896-grover-search-algorithm-standard-and-enhanced-versions

## Biography

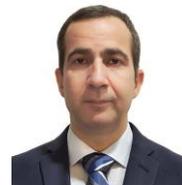

Ismael Abdulrahman received his PhD in Electrical and Computer Engineering from Tennessee Technological University, USA, in 2019. Currently, he serves as an assistant professor at Erbil Polytechnic University, where he instructs graduate and undergraduate courses including quantum computing, advanced mathematics, and electrical and electronic courses. His academic passions include quantum computing, machine learning, optimization, and mathematical modeling.